# Screening of common synthetic polymers for depolymerization by subcritical hydrothermal liquefaction

Juliano Souza dos Passos[1], Marianne Glasius[2], Patrick Biller[1]*
1. Biological and Chemical Engineering, Aarhus University, Hangøvej 2, DK-8200 Aarhus N, Denmark
2. Department of Chemistry, Aarhus University, Langelandsgade 140, DK-8000 C, Denmark

*Corresponding author: pbiller@eng.au.dk

**Abstract.** Hydrothermal liquefaction could potentially utilize mixed plastic wastes for sustainable biocrude production, however the fate of plastics under HTL is largely unexplored for the same reaction conditions. In this study, we evaluate how synthetic waste polymers can be depolymerized to bio-crude or platform chemicals using HTL at typical conditions expected in future commercial applications with and without alkali catalyst (potassium hydroxide). We evaluate different characteristics for HTL processing of poly-acrylonitrile-butadiene-styrene (ABS), Bisphenol-A Epoxy-resin, high-density polyethylene (HDPE), low density PE (LDPE), polyamide 6 (PA6), polyamide 66 (PA66), polyethylene terephthalate (PET), polycarbonate (PC), polypropylene (PP), polystyrene (PS) and polyurethane (PUR) at 350 ºC and 20 minutes residence time. Polyolefins and PS showed little depolymerization due to lack of reactive sites for hydrolysis. HTL of PC and Epoxy yielded predominantly bisphenol-A in oil fraction and phenols in aqueous phase. PA6 and PA66 yielded one of its monomers caprolactam and a range of platform chemicals in the aqueous phase. PET produces both original monomers. PUR yields a complex oil containing similar molecules to its monomers and longer hydrocarbons. Our results show how HTL can depolymerizes several different synthetic polymers and highlights which of those are the most attractive or are unsuitable for subcritical processing.

## 1. INTRODUCTION

Our modern society relies on an unsustainable linear production of goods, entailing the extraction of natural resources, refining and production of consumer goods and commodities followed by final disposal. The unavoidable end-of-life products of such processes are either mechanically recycled, disposed of in landfills or undergo combustion if suitable.(Korhonen et al., 2018) The latter option, despite reducing solid residues in landfills, releases $CO_2$ in the atmosphere while recovering heat as the lowest value commodity.

Circular economy is an alternative mode of production(Ghosh and Agamuthu, 2018; Korhonen et al., 2018) in which industry considers waste streams generated by society as its own source of raw materials. Innumerous implementation challenges are present here, e.g. mixed materials where each substance should be diverted to a different sector; combined goods, which include synthetic polymers, inorganic salts and metals fused together. Scientists around the world are trying to solve this problem by creating more efficient and less costly innovative solutions. The hydrothermal liquefaction (HTL) technology is a promising alternative to fulfil these requirements, as it is highly flexible in dealing with both pure waste streams and mixed ones (Biller et al., 2018).

The HTL concept has been investigated since as early as in 1982(Coorporation, 1982), with different approaches being introduced over the years(Foudriaan and Peferoen, 1990), however early efforts aimed at substituting crude oil with bio-based feedstock for political reasons instead of climate concerns. Nowadays, the latter reason has encouraged many research groups to investigate the HTL processing efficiency and its technical feasibility in depth(Anastasakis et al., 2018; Skaggs et al., 2018). HTL has proven so far to be a technology able to with a wide range of organic waste feedstocks(Anastasakis et al., 2018; Castello and Pedersen, 2018; Skaggs et al., 2018) and synthetic polymers(Pedersen and Conti, 2017) for the production of a bio-crude which can readily be upgraded to transportation fuels(Castello et al., 2019). To this date HTL research has primarily focused on biofuel production from biomass and wastes while the utilization of waste polymers has received little attention.

Single synthetic polymers subcritical HTL has been reported for specific materials in several publications, including reports on HTL of: high-impact polystyrene, poly-acrylonitrile-butadiene-styrene (ABS), polycarbonate and polyamide 6(Iwaya et al., 2006; Zhao et al., 2018a); epoxy printed circuit boards(Yildirir et al., 2015); polyethylene naphthalate and terephthalate(Arai et al., 2010; Zenda and Funazukuri, 2008); polystyrene-butadiene(Park et al., 2001); polyurethane(Dai et al., 2002). These studies show that monomers, other valuable chemical compounds or an oil product may be recovered using HTL. However, comparable data on the depolymerization mechanisms of these polymers is unavailable, as different reactor setups, residence times, heating rates, temperatures and pressures reported in literature make it difficult to assess if the current HTL technologies used for biomasses and organic waste(Anastasakis et al., 2018; Biller et al., 2018, 2016) are suited for synthetic polymers as well. In general, current subcritical HTL technology relies on fast heating rate reactors with moderate residence time (15-20 minutes), working close to water saturation pressure in the range of 300-360 ºC and, depending on feedstock, applying alkali catalysis (e.g. $K_2CO_3$, KOH).

The present study investigates how HTL can be applied as a generic valorization process for synthetic polymers using the same conditions as in modern biomass liquefaction. We hypothesize that HTL of polymer waste has positive and negative aspects that must be unraveled for each synthetic material in order to assess if they can be included in a combined HTL waste treatment processes for e.g. mixed municipal waste or as a standalone technology for chemical plastic recycling. Batch experiments were conducted to evaluate how the most used synthetic polymers behave when processed using fast heating rate reactors and short residence times. We give a comprehensive overview of the fate of the most common waste polymers at a given condition to allow a fully comparative assessment for future implementation in the circular economy.

2. MATERIALS

A total of 12 different commercial polymers – poly-acrylonitrile-butadiene-styrene (ABS), Bisphenol-A based Epoxy resin, high density polyethylene (HDPE), low density polyethylene (LDPE), non-colored plastic cable ties of polyamide 6 (PA6), Sigma-Aldrich polyamide 6/6 (PA66), polyethylene terephthalate plastic bottles (PET), polycarbonate (PC), polypropylene cups (PP), polystyrene cups (PS) and polyurethane foam (PUR) – were milled using a Polymix® PX-MFC 90D knife mill equipped with a 2 mm sieve before the HTL procedure. Sigma-Aldrich Poly(vinyl chloride) (PVC) was used as acquired (powder).

3. METHODS

## 3.1. HTL procedure

Reactions were performed using custom made Swagelok© bomb-type 20 mL reactors following already described procedures(Biller et al., 2016). Two sets of experiments were conducted in duplicates, one with pure water and one with alkali catalyst (KOH). In each experiment, 0.50 g of polymer was added to the reactor with 8.5 g of water or alkali aqueous solution (17.2 g/L). The feed dry matter concentration of the experiments (5,6%) is lower than commonly applied in continuous systems (typically 15-20 %) (Castello and Pedersen, 2018). Such dry matter content was chosen to standardize loadings for all polymers tested, as some polymers were difficult to fill into the limited reactor space due to their low density. Reactors were sealed and submerged into a pre-heated fluidized sand bath for 20 minutes at 350 ºC, which results in a heating time of 4 minutes (average heating rate of 82 ºC min$^{-1}$). This heating rate approximates the heating profile of a modern continuous HTL plant with integrated heat recovery, where a similar heating rate of 75-100 °C/min was applied and a total reaction time of approximately 14 minutes.(Anastasakis et al., 2018) The reactors were then quenched in a water bath, cleaned and weighted. The gas produced during the HTL reaction was carefully vented and the reactor re-weighted to determine the mass of gas generated. The aqueous phase (AP) was transferred to a 15 mL centrifuge tube and centrifuged for 5 min at 4000 rpm. To recover the oil phase, 30 mL of methanol was used to wash remaining solids both in the reactor and in the AP centrifuge tube. After filtering the methanol, an aliquot of 1 mL was withdrawn for gas chromatography/mass spectrometry (GC/MS) analysis, while the remaining liquid was evaporated overnight at 35 ºC in a convection oven to determine the oil weight. Solid residues were dried at 105 ºC overnight in a convection oven and AP mass was determined by difference. All yields were based on the mass of polymer initially used.

## 3.2. Elemental analysis

An Elementar vario Macro Cube elemental analyser (Langenselbold, Germany) was used to determine the CHNS content of all raw materials, solid residues and oil products in duplicate, average values are reported. The combustion chamber was operated at 1150 ºC and the reduction tube at 850 ºC, with a helium flow of 600 mLmin$^{-1}$.

### 3.3. ATR-FTIR

A Bruker Alpha Platinum Attenuated Total Reflectance Fourier-transform infrared spectroscopy (ATR FTIR) spectrometer was used to collect 24 spectra from 4000 to 400 cm$^{-1}$ with resolution of 2 cm$^{-1}$. The ATR crystal was cleaned using 96% ethanol and baseline signal was checked between measurements. Solid samples were compressed against the diamond crystal and liquid/grease samples were rubbed on top for measurements.

### 3.4. GC/MS

Analysis were performed using an Agilent 7890B GC coupled to a quadrupole mass filter MS (Agilent, 5977A). For oil analysis, 1.0 µL of the aliquot retrieved from sample work-up was directly injected (inlet temperature of 280 °C, split ratio 20:1, helium flow 1 mL.min$^{-1}$) after internal standard addition (4-bromotoluene) on a VF-5ms column (64.9 m x 0.25 mm x 0.25 µm). The following GC oven temperature program was performed: 60 ºC hold for 2 minutes; ramp to 200 ºC (5 ºC.min$^{-1}$); ramp to 320 ºC (20 ºC.min$^{-1}$); hold for 5 minutes. Compounds were identified with authentic standards, NIST17 mass spectra library or based on literature references. Quantification was conducted using an 8-point calibration curve for selected compounds.

For AP analysis, the methyl chloroformate derivatization method was used (see reference(Madsen et al., 2016) for full details). Catalysts and methyl chloroformate were added to the AP to methylate water soluble compounds. Later, chloroform containing internal standard (4-bromotoluene) was used to extract these compounds to an organic phase, which was then analyzed using GC/MS.

### 4. RESULTS AND DISCUSSION

Figure 1a and 1b show the mass balance for all polymers tested. The polymers are divided into (a) polymers which are prone to HTL conversion and (b) polymers which are not prone to conversion. For the polymers which do convert (Figure 1a) the amount of material fractionating to the aqueous phase (AP) constitutes an important fraction of the mass balance for most polymers, particularly when alkali catalyst is present. Higher reaction rates for hydrolysis depolymerization mechanisms have previously been reported for PET in presence of alkali (Wan et al., 2001), which is confirmed in the current study for other polymers, as shown in Figure 1a by markedly increased polymer conversions for all samples apart from PVC. Alkali reactions are responsible for breaking –O– and –N– structures into alcohols, carboxylic acids, amines or amides.(Singh and Sharma, 2008) In presence of alkali, hydrolysis reactions yield organic salts as well, which have higher water solubility.(Singh and Sharma, 2008) Hydrolysis reactions depend on a common characteristic of most polymers grouped into Figure 1a: the presence of heteroatoms in their backbone structures.

The polymers subjected to HTL (see Table S1 for chemical structures) were grouped into the following topics for discussion according to important observations and characteristics during HTL processing:

I) HDPE, LDPE, PP and PS: where the absence of heteroatoms in polymeric structures prevents HTL depolymerization mechanisms (Figure 1b);

II) ABS: a polymeric structure without backbone heteroatoms that shows increased depolymerization in alkali conditions;

III) Epoxy and PC: comparison of two BPA-based polymers – one thermoset and one thermoplastic – evaluating how cross-linked polymers decompose differently;

IV) PET: a polymer yielding a large amount of solid products;

V) PA6 and PA66: where crystalline and amorphous polymer phases play a role on depolymerization mechanisms;

VI) PUR: a thermosetting polymer containing both N and O heteroatoms;

VII) PVC: process occurs in heavily acidic HTL conditions due to halogenated raw materials.

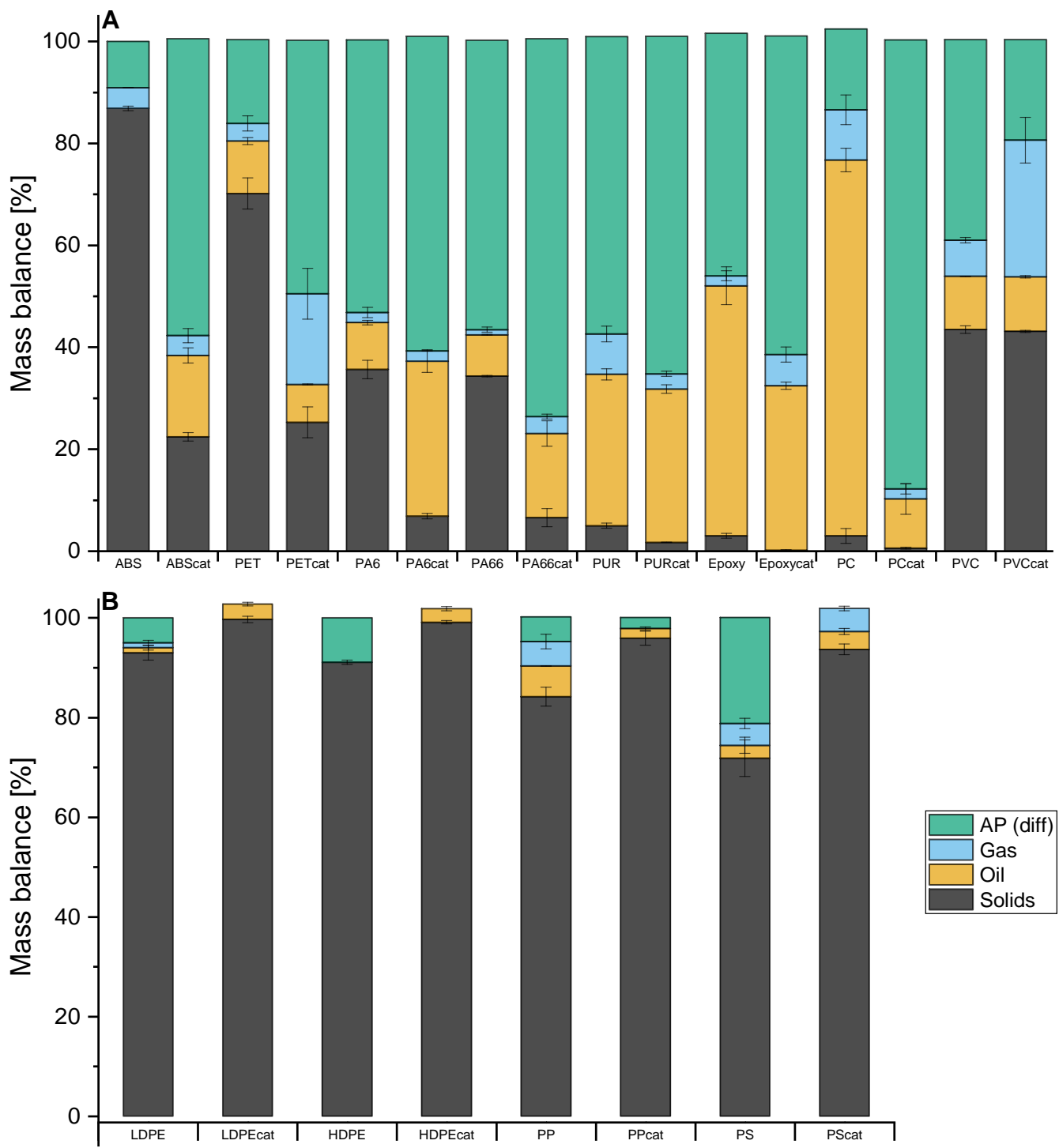

Figure 1. Mass balances for polymers after HTL (cat = KOH catalyzed HTL). (a) Polymers prone to HTL conversion. (b) Polymers not prone to HTL treatment

## 4.1. HDPE, LDPE, PP and PS

All of these polyolefins showed >90% solid residue yields after HTL processing (see Figure 1b). Additionally, when processing in the presence of alkali, solid residues increase for all four polymers. The main reason is that no heteroatoms or reactive sites are found in these materials, thus thermal cracking is the preferred depolymerization mechanism (Singh and Sharma, 2008). PE has been reported to decompose to oil products under supercritical water (450-480 °C), above its thermal cracking temperature (commonly around 435 °C) (Hai-feng et al., 2007; Su et al., 2004). When compared to pyrolysis, the presence of supercritical water prevents coke formation while increasing oil yield (Moriya and Enomoto, 1999). Subcritical water LDPE HTL, with hot release and condensation of vapours, has been reported once by Wong et al. (2005), showing an organic liquid yield of 28.42% at 300 °C with residence time of 120 min.(Wong et al., 2016) As the temperature of 350 °C used in our experiments, is lower than the thermal cracking temperature of PEs in general, the reaction rate is too low for the time used in our experiments (20 min), leaving polymeric chains in solid state.

FTIR analysis of solid residues showed that some are partially oxidized (see Figure 2S A-D). HDPE and LDPE non-catalyzed HTL resulted in oxidized materials containing =O bonds (1712 cm$^{-1}$). Catalyzed reactions yielded a non-oxidized HDPE and a different type of oxidized LDPE, containing also C=C–O (1648-1626, 835 cm$^{-1}$) and C–O (1400 cm$^{-1}$-1256 cm$^{-1}$) groups. PP solid products of non-catalyzed HTL also show C–O (1285 cm$^{-1}$), C=O (1680, 1714 cm$^{-1}$) and C=C–H (934, 841, 732 cm$^{-1}$) groups, while catalyzed HTL of PP did not change the material according to the FTIR characterization.

HTL of PP has only been reported for long reaction times and low heating rate batch experiments. (Zhao et al., 2018b) For 1h reaction time at 350 °C, 65% solid residue yield was reported, with oil products containing cyclic alkanes and aliphatic alkenes. Similar products are also present in the small oil fraction recovered in our experiments, although at very low levels (see chromatograms in Figure S5).

PS did not show any structural changes by HTL using FTIR analysis. A similar polymer to PS, a co-polymer of styrene with butadiene, has been reported to decompose in the presence of water at 350 °C with depolymerization as high as 61% for a reaction time of 60 min.(Park et al., 2001) Near supercritical HTL (370 °C) was also reported to decompose PS into valuable chemicals in short residence times(Kwak et al., 2005), however, supercritical PS HTL(Kwak et al., 2005; Park et al., 2001) was recently found to have the most efficient condition at 490 °C with carbon liquefaction efficiency as high as 80%.(Bai et al., 2019) In the present study, the aromatic-containing polymer PS has shown to be stable under subcritical conditions.

The minor changes of H/C and O/C ratios change during HTL, presented in Figure 2, indicate that HDPE, LDPE, PP and PS experience no measurable change in their chemical structures. The absence of heteroatoms in the polymer backbone structure prevents all depolymerization mechanisms that involve reactive water in the investigated HTL conditions. Hence it can be concluded that the inclusion of these polyolefin wastes should be avoided in HTL feedstocks for biocrude production in any future mixed waste application at subcritical conditions and supercritical conditions are favorable. A pyrolysis approach where higher temperatures, above the thermal cracking temperature, is employed or supercritical liquefaction are required conditions for the chemical recycling of these materials.

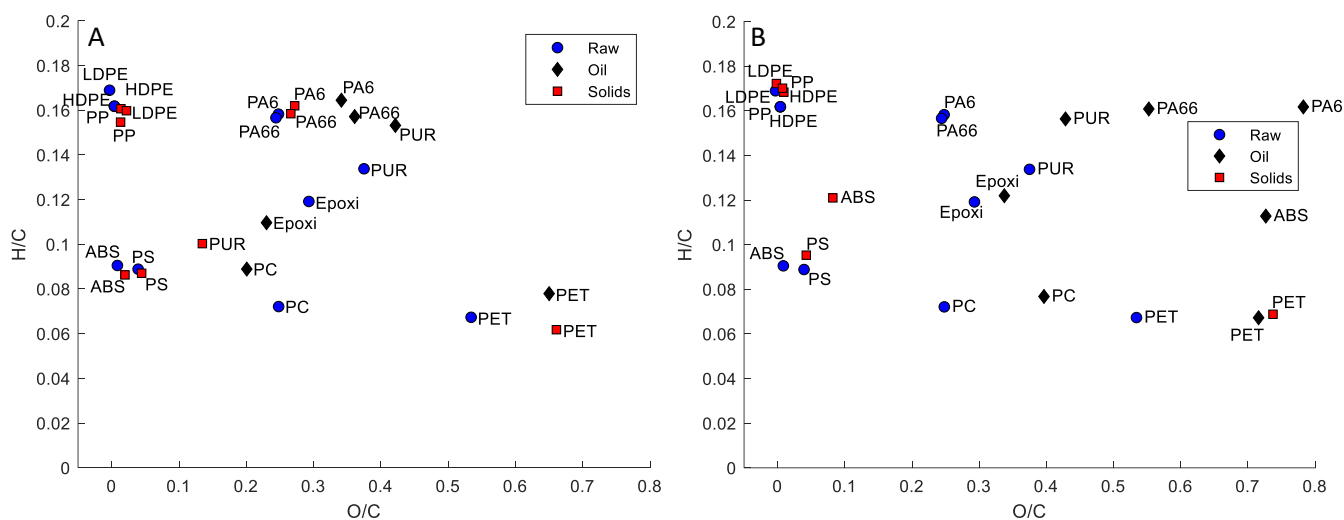

Figure 2. Van Krevelen diagram of the different phases – without (A) and with catalyst (B)

### 4.2. ABS

Even though the chemical structure of this polymer does not contain backbone heteroatoms to favor hydrolysis depolymerization reactions, ABS shows a positive effect on decreasing solid residues with alkali catalysis (Figure 1a). This result is a consequence of the other reactive sites present in ABS, nitrile side group and the backbone double bond, becoming prone to depolymerization in alkali media.

The GC/MS analysis shows (Figure S4 and S5) that aromatic ≡N molecules are present in non-catalyzed AP products, however, when alkali catalyst is present, only $-NH_2$ and $-NH-$ groups are detected. This indicates that alkali conditions stimulate depolymerization reactions by forcing nitrile hydrolysis. Principally, nitrile groups hydrolyze in presence of hot compressed water(Izzo et al., 1999) generating $NH_3$ and carboxylic acids with an amide intermediate (-CONH$_2$).(Izzo et al., 1997) Potassium hydroxide has been reported to catalyze this reaction at low temperatures(Sanli, 1990) (20-80 °C). The same effect is observed here, by generating ammonia under HTL conditions, the depolymerization rate of the materials is increased.

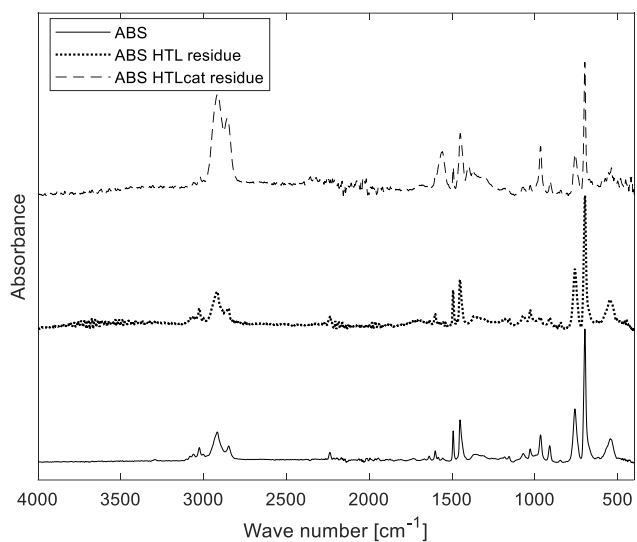

Figure 3 – FTIR of ABS solid residues in comparison to raw material.

Besides the ammonia generation path, as double bonds are present in butadiene units, direct addition of the amide intermediate(Sanli, 1990) to the alkene section(Mu et al., 2008) of butadiene units are possible. According to the FTIR data in Figure 3, HTL without catalyst only changes the polymeric structure regarding the –C=C– bonds, which is evident due to the disappearance of sharp peaks at 967, 910 and 1638 cm$^{-1}$, meaning that such double bonds are either hydrogenated or substituted by rearrangement. On the other hand, the spectra found for the solid residue of catalyzed HTL suggests a rearranged structure containing more aliphatic sections and a –C–N=C– bond (1158 cm$^{-1}$), a derivative of amide addition to butadiene units. Despite the evidence for a –C–N=C– bond, the elemental analysis depicted in Table S3 shows that the solid residue only contains 0.31% of N, a much lower value compared to the original content of 5.13%.

Hence, KOH catalyzes nitrile hydrolysis, which generates reactive ammonia that interacts easily with butadiene and styrene units for depolymerization, generating oxygenated compounds that are more soluble in water. For the case of ABS, hydrolysis is not the main pathway for depolymerization, but the chain reaction started by a side group (–C≡N). The products generated by these reactions are hydrophilic – probably composed of oligomers due to the absence of small structures by GC/MS analysis – and only small amount of oil being generated.

### 4.3. Epoxy and PC

The two polymers have related chemical structures as epoxy is a thermoset composed of bis-phenol-A diglycidyl ester (BPA-DGE) and butandioldiglycidylether (BODGE) cross-linked by poly(oxypropylen)diamine and aminoethyl-3,5,5-trimethylcyclohexylamine, while PC is a thermoplastic of BPA. As both structures contain backbone oxygen heteroatoms, hydrolysis is the preferred depolymerization path. It is important to remark that one of the crosslinking reagents of Epoxy has significant nitrogen content (see Table S1 and S2), which reacts into $NH_3$ at HTL conditions, catalyzing depolymerization reactions.

Alkali effect on both reactions is qualitatively similar (Figure 1a), as it increases the water-soluble products in different degrees. AP yield for PC increases from 15.9% (no-cat) to 88.2% (cat), while from 47.6% (no-cat) to 62.5% (cat) for Epoxy. The products obtained in both non-catalytic processes are typically alcohols, which can further react in presence of KOH into carboxylic acids or smaller alcohols, having higher water solubility.

The nitrogen content in Epoxy can explain the higher water-soluble products yield for the material in comparison to PC under non-catalytic HTL, as both oil-yielded reaction products are very similar (Figure 4). Also in Figure 4, it is possible to observe that even though Epoxy is a heavily cross-linked polymer, it yielded very similar compounds as its thermoplastic counter-part (PC). Aqueous phase GC/MS analysis shows that phenol is present in both catalytic and non- catalytic HTL reactions. However, the phenol yield is more prominent when KOH is present (see "Chemical recovery of platform chemicals" for further discussion).

For both Epoxy and PC, very low amounts of solids were obtained, showing that despite high levels of crosslinking in Epoxy, HTL conditions promoted severe depolymerization reactions as in PC. This fact corroborates that chemical structure (i.e. backbone heteroatoms) is more important than crosslinking for HTL treatment.

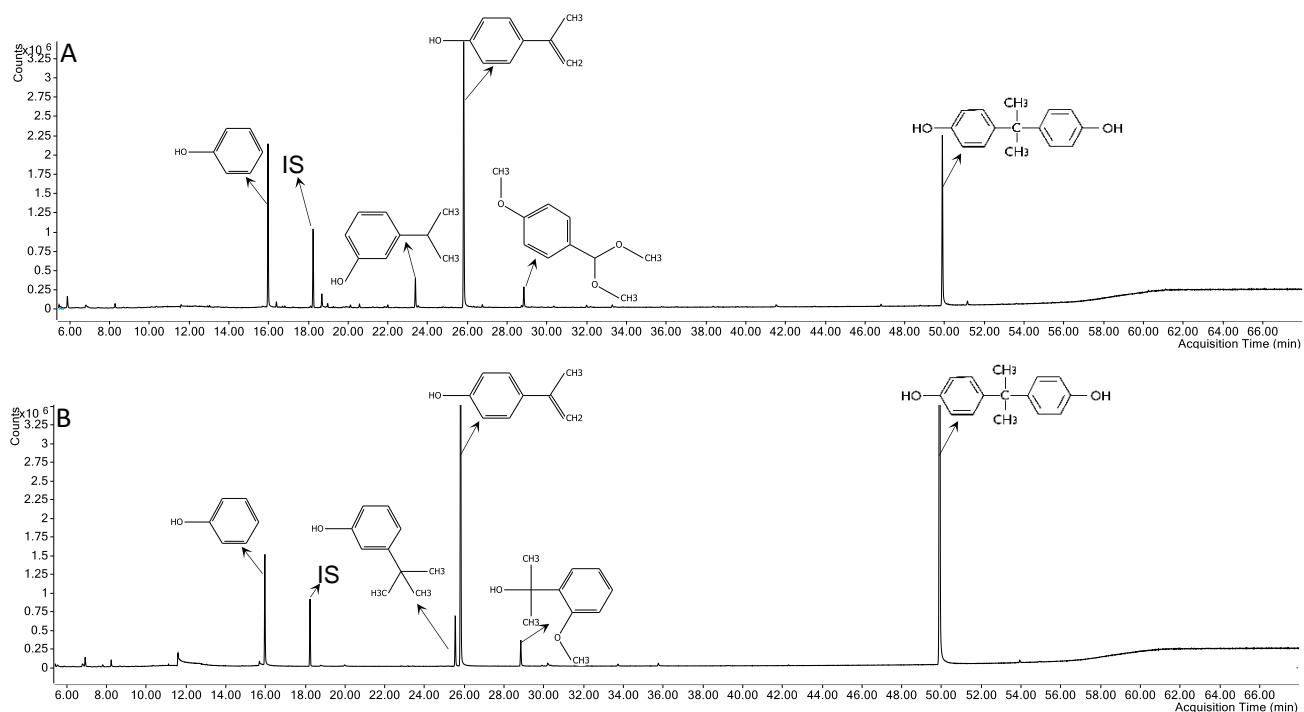

Figure 4 – Epoxy (A) and PC (B) oil products GC/MS. (IS = internal standard)

## 4.4. PET

This condensation polymer differs greatly in behavior to others, as one of its monomers – terephthalic acid (TA) – is an insoluble solid in both water and alcohols. Figure 5 presents the FTIR spectra that shows clearly a high-purity TA as result of HTL procedure given the FTIR reference match. In this case, Figure 1 shows an increase of AP for catalytic HTL, which demonstrates that alkali catalysis converts TA into its salts, increasing its solubility in water. Besides, the higher gas yield for catalytic HTL shows that decarboxylation reactions are favored, generating more $CO_2$ at the studied temperatures. This can be observed by the higher gas amount for catalytic HTL of PET in Figure 1a and the GCMS findings in Figure S6 and Figure S7, which show direct decarboxylation products of TA.

Again, the presence of oxygen as heteroatoms contributes greatly to depolymerization reactions – mainly hydrolysis in this case.(Wan et al., 2001) PET depolymerization reactions have been studied before at lower temperatures (120-160 °C) than reported here, though also in presence of KOH.(Wan et al., 2001) Higher temperatures (190-307 °C) were also tested in a different study by Zenda & Funazukuri (2008)(Zenda and Funazukuri, 2008) in presence of other alkaline catalysts (ammonia and NaOH) yielding similar results.

Both cases achieve very high terephthalic acid yields, as also shown in this study. Figure 1a shows that PET HTL in presence of KOH exhibits a significantly higher gas yield, indicating catalytic decarboxylation of TA occurs at the temperature tested. For the case of catalytic PET HTL, the oil and water yields contain benzoic acid, a direct product of TA decarboxylation.

The presence of a valuable solid product raises an issue on PET HTL processing: how to deal with the solid streams in an application context. For the case of pure PET HTL, direct use of the solid stream is an advantage, however, if PET co-processing through HTL is desired, such valuable solid will be mixed with different byproducts. In this case, extraction and purification of the solid stream has to be considered as an essential step.

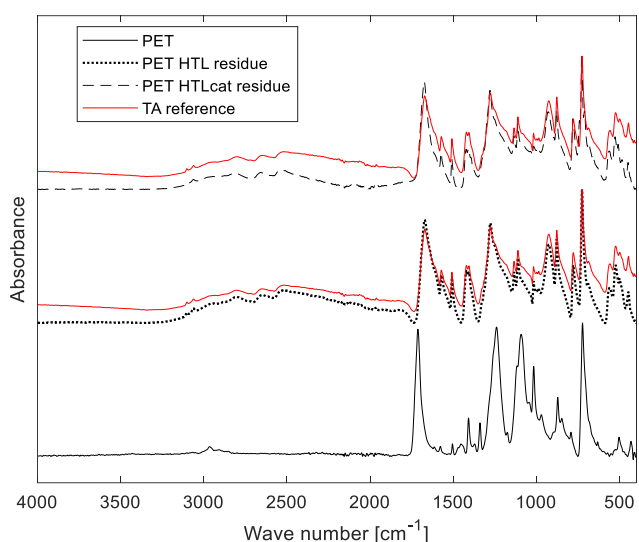

Figure 5. FTIR of PET solid residues in comparison to raw material including TA reference.

## 4.5. PA6 and PA66

Figure 1a depicts a similar mass balance for PA6 and PA66 with slightly lower oil yield for catalyzed HTL of PA66 in comparison to catalyzed HTL of PA6, in both cases the addition of alkali approximately doubled the oil yield. For both PA6 and PA66, monomers, dimers and some variations of these compounds are the products identified in GC/MS analysis of the AP (see Figure S8 and S9). Similar results were previously reported for non-catalyzed HTL (Iwaya et al., 2006), but surprisingly little effect on product distribution could be observed when KOH was present as catalyst despite the increase in oil yield.

The O/C ratio of PA6 oil products of catalytic HTL was higher than of the corresponding PA66 sample. Both were significantly different from the non-catalytic HTL products (see Figure 2). The oxygen migration to the oil phase in PA6 when comparing non- and catalytic HTL is probably caused by a higher hydrolysis rate, which decomposes caprolactam (the main monomer) into further hydrolyzed products, which contain more oxygen.

Analysis of the FTIR spectra in Figure 6 confirms that the amorphous phase of PA6 and PA66 (1145 cm$^{-1}$) (McKeen and McKeen, 2012) are depolymerized under non-catalytic HTL. For PA6, when alkali was used for HTL, the spectra show a significant difference between peaks at 665 cm$^{-1}$ and 580 cm$^{-1}$, showing that α phase is the dominant type of PA present. i.e., γ phase was more prone for depolymerization, which is expected as α phases are the most stable for this type of polymer.(McKeen and McKeen, 2012) Here, the alkali again promotes higher rates of hydrolysis, however the crystalline structure dictates the depolymerization path.

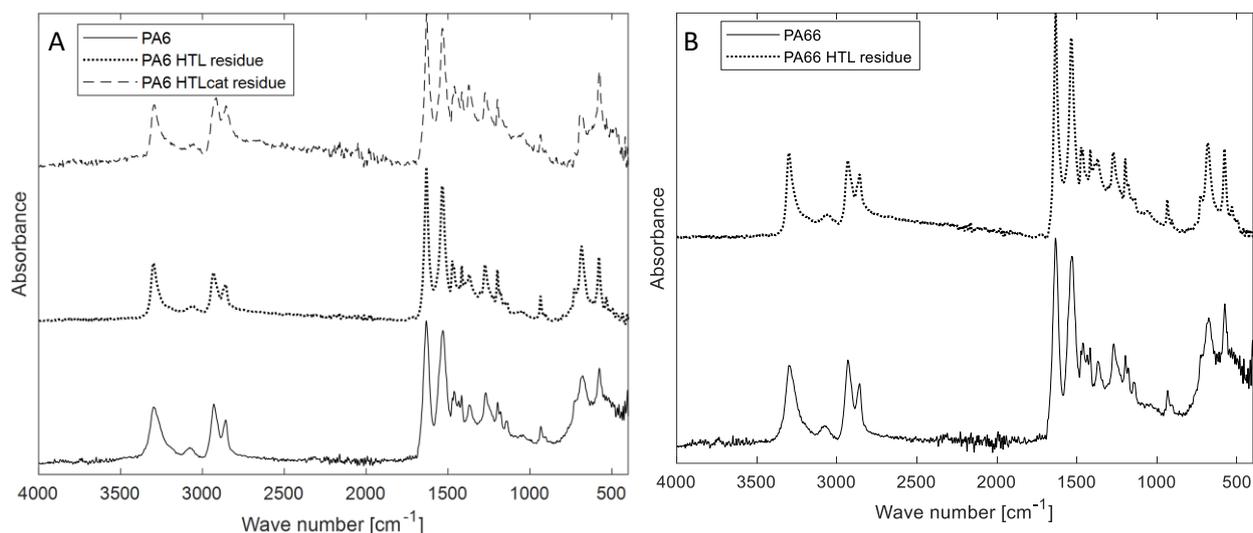

Figure 6. FTIR PA6 (A) and PA66 (B) – Original polymer and solid residues of HTL with and without catalyst (KOH)

### 4.6. PUR

Figure 1a and Figure 2 show that PUR does not exhibit major differences in mass balance nor in H/C and O/C ratios for catalytic HTL compared to non-catalytic HTL. The catalyst does have an effect of increasing quantified compounds, an observation which is discussed later. The lack of catalytic effect on H/C, O/C ratios for this polymer suggests that the amount of N present was already enough to generate sufficient in situ NH$_3$, promoting hydrolysis by itself. This indicates that PUR is prone to catalyze degradation of other polymers under co-liquefaction, which opens up opportunities for positive synergistic processes for polymer or biomass combinations. PUR hydrolysis has been reported to yield mainly diamino toluene using long retention times and slow heating rate reactors.(Dai et al., 2002) Our study shows that for fast heating rate and short residence times, oligomers are the main products, with a clear phase separation. Figure S10 presents the FTIR spectra of PUR oil, highlighting that secondary amines (3400-3200 cm$^{-1}$), N–C=O (1734 cm$^{-1}$) and C–N (1222 cm$^{-1}$) characteristic bands are not present in both catalytic and non-catalytic HTL oil product. It also highlights that the C–O–C band at 1089 cm$^{-1}$ is present in all samples, indicating that the oil products also have ether groups, which corroborates that oligomers are the main components in this product stream.

Figure S11 shows that N–H (3300 cm$^{-1}$) bonds are not present in the solid residues of non-catalyzed HTL, however aromatics are much more prominent (indicated by 700, 757, 1452 and 1495 cm$^{-1}$), which suggests that rearrangement of the polymer happened to yield a more stable structure. Even though aromatics are evident in these solid residues, they are present together with O heteroatoms (1089 cm$^{-1}$).

The AP yield increased from non-catalytic to catalytic HTL due to the reduction in gas yield, which indicates less decarboxylation and thus, more carbon retention in liquid products. The composition of this stream did not change significantly, as can be seen from the chromatogram in Figure S12. The products identified in this phase are aromatics and polyaromatics containing O and N heteroatoms, however low match factors from the library search prevents reliable identification, due to characteristic branched compounds being identified.

### 4.7. PVC

The only halogenated polymer exhibited pronounced charring reactions rather than liquefaction reactions due to the acidic conditions. Addition of alkali resulted in no considerable differences in oil and solid yields (Figure 1a), however, a major difference was measured on gas and AP yields. The higher gas yield in alkali conditions indicates that a greater portion of chlorine present was converted into $Cl_2$, reducing the mass transfer to the AP. Despite this, the solid residues from catalyzed HTL have a lower carbon content (Table S1) compared to those from non-catalytic HTL, indicating that the catalyst lower the polymer-solid residue dechlorination effect.

HTL processing of PVC showed that it is capable of recovering a solid residue with much lower chlorine content, which could be an approach for solving incineration issues related to such contaminants. However, PVC should be avoided in HTL co-processing as acidification is undesired and unavoidably results in carbonization reactions.

## 5. ANALYSIS OF DEPOLYMERIZATION PRODUCTS IN OIL PHASE

Figure 7 shows the quantified compounds present in oil products for PC, Epoxy, PUR and PA66 HTL with and without catalyst. Of these, PC is the polymer with highest oil yield, which is composed of relatively few compounds, in particular its monomer, BPA, contributing 44.5 wt.% of the total oil yield. This means the total chemical recycling efficiency of PC to BPA is 32.8 wt.%, to p-isopropenyl phenol is 20.2% and to phenol is 5.2%. Part of the Epoxy oil yield is also composed of the same products as PC, however in lower concentrations, yielding a total chemical recycling efficiency of 5.6 % to p-isopropenyl phenol, 4.1 % to phenol and 2.6 % to BPA. Epoxy is a crosslinked thermoset and thus has a branched structure which leads to more complex depolymerization mechanisms and, by extension, more complex products. The presence of ether-like sections, generated by epoxy bonds in the original structure (see Table S1) leaves BPA attached to branched alcohols, which contributes to the non quantified compounds (NCQ).

PUR has been reported to hydrolyze in hot and compressed water (Brunner, 2014), however at the conditions tested, it seems that only partial decomposition is achieved. As shown in Figure 7, the PUR oil only exhibits a minor amount of small organic compounds, suggesting that the NQC in this case are oligomers of the original structure. Full conversion of PUR can only generate water soluble compounds (Brunner, 2014), which would make it difficult for recovery of valuable compounds. Alkaline catalysis increases the concentration of identified oil products, which indicates that smaller compounds are produced. Thus, the HTL procedure suggested here could be an option for high carbon recovery of waste PUR with the advantage of having a phase-separated product that can be further purified into an oil product or even individual chemicals.

Figure 7 also shows that hydrazide benzoic acid is the main quantified oil product from PA66. When catalyzed with alkali, the products of PA66 HTL have higher concentrations of oxygen (Table S2 and S3) and lower of carbon, representing a more hydrolyzed product with lower value.

Besides the oil phases, AP products of polymer HTL are significant often making up over 50% of the total mass yield. Their low concentrations represent an issue for recovery or valorization of chemicals and energy recovery via e.g. hydrothermal gasification is a more likely route.(Elliott, 2011) Out of the polymers tested, the most favorable AP products are the ones resulting from PA6, PA66, Epoxy and PC HTL if chemical recovery from non mixed wastes is the objective. In the two former cases, the AP is reported(Iwaya et al., 2006) to be composed mostly by ε-caprolactam (approx. 80% is converted of the original polymer), which could potentially be re-polymerized into new bulk polymers. For the two latter, phenol is the most abundant product in the alkali-catalyzed HTL of AP, constituting 5.3; 7.6; 29.9 and 11.0% of the mass present for AP of Epoxy, Epoxycat, PC and PCcat, respectively.

For PET HTL, the solids recovered were the main product. They are composed of high purity TA, which is the monomer of PET. The decarboxylation observed when PET was processed with catalyst also yields useful platform chemicals that could be further processed in a HTL context where only PET is used as the feedstock. However we consider that this would be a downgrading of such a valuable monomer, as if PET is processed in its pure form (non-catalysed), the solid TA is recovered immediately. If co-processing is desired to eliminate the need of pure streams for HTL, a solid product represents a challenge for further separation.

The co-processing of different synthetic polymers in HTL for platform chemical recovery can offer advantages and disadvantages. In one hand, co-processing of selected polymers can offer synergies on specific recoveries, e.g. PC and Epoxy may be co-processed and the final oil phase will contain most of the BPA recovered for both, meanwhile the AP will comprise phenolics. On the other hand, the generation of certain products can interfere in the recovery of others, e.g. ethylene glycol generated in PET AP products can increase the solubility of organic compounds in other polymer's oil phase, which would increase separation costs. As valuable compounds are shown here, simple separation processes may be considered (liquid-liquid extraction, evaporation or distillation), however with the increase of mixture complexity in co-processing scenarios, increased cost may overcome such approaches' feasibility.

The processing of synthetic polymer waste in high thermal efficiency HTL reactors(Anastasakis et al., 2018) thus yields some ready-to-use platform chemicals which could be recovered if pure polymer streams are used, which can represent an advantage in comparison to other recycling strategies, such as mechanical recycling, as this approach is not limited by a number of recycling cycles. Overall, the results however also show that generally the composition of products is complex and would become even more if unsorted streams of synthetic wastes are used or mixed with biomass wastes. In this context it can be concluded that some of the polymers are beneficial in such approaches, as polyamides, PUR, epoxy and PC, yielding additional oil products, while others are uncertain, as ABS and PVC and their co-processing deserves additional attention. The polyolefins in general should be avoided, adding solid residue to the product mixture and little conversion at the subcritical conditions investigated here.

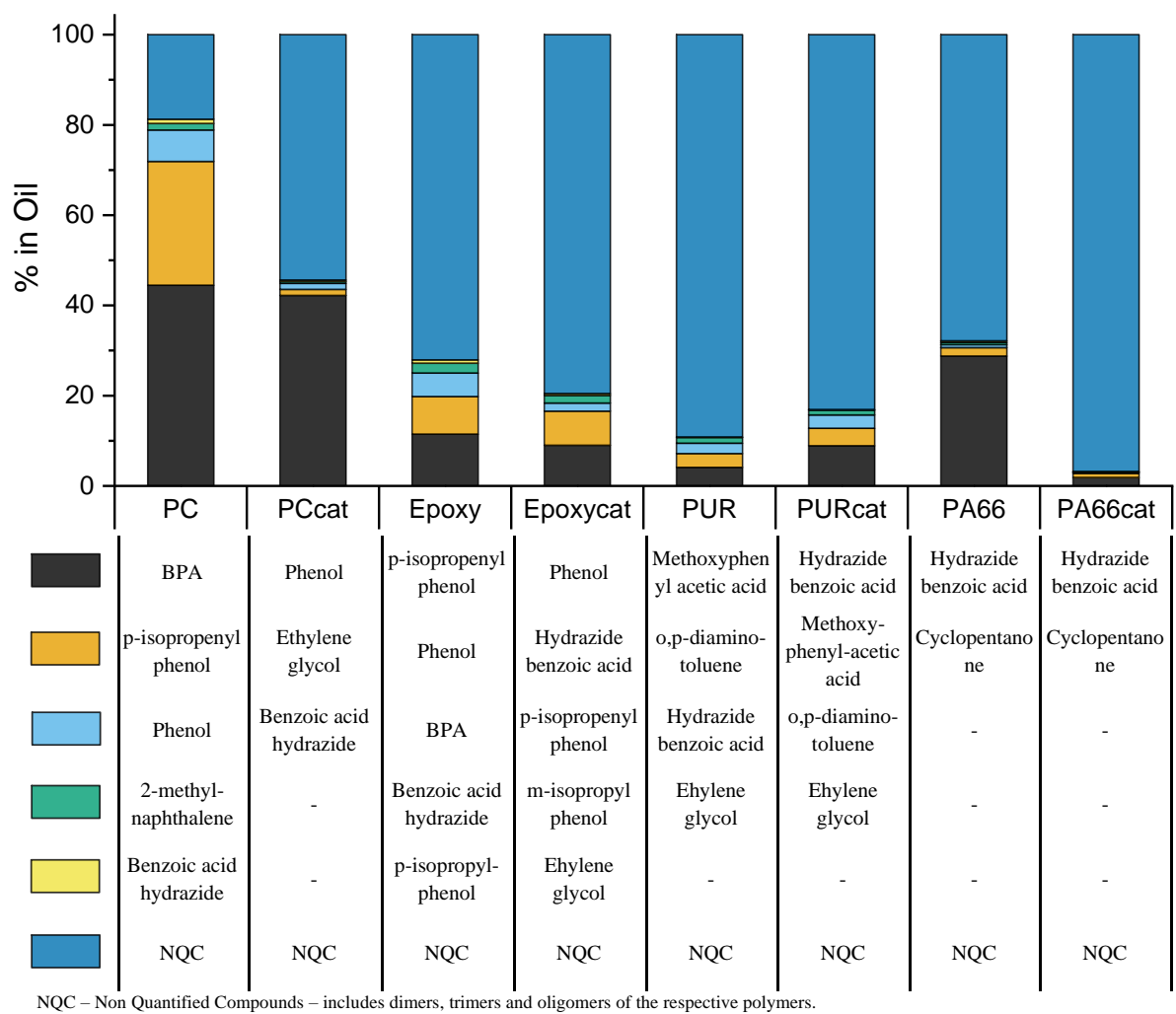

NQC – Non Quantified Compounds – includes dimers, trimers and oligomers of the respective polymers.

Figure 7. Oil phase mass composition by quantitative GC/MS analysis.

## 6. CONCLUSION.

Subcritical HTL processing of synthetic polymers is a promising approach for certain types of polymers chemical recycling. Products identified and quantified for the case of suitable feedstock are valuable chemicals, thus this approach may contribute for a circular economy of the sector. The HTL process is intrinsically dependent on reactive sites for hydrolysis in the original chemical structures of the polymers and does not occur in absence of those. This creates a challenge for subcritical processing of polyolefins and polystyrene. On the other hand, for materials that present such reactive sites, a major part of the mass processed result in the AP, which creates difficulties in down-stream valorization. The alkali environment provided by the catalyst used in this study showed to favor a greater hydrolysis rate, generally decreasing solid residues and increasing AP yields.

For all other polymers evaluated in this study, particular process characteristics can be highlighted as follows:

ABS: alkali significantly increases depolymerization rates, however products are concentrated in the AP. The oil product generated is composed of oligomers of the original structure and requires purification and upgrading to recover platform chemicals;

Epoxy and PC: BPA and derived compounds are present in the oil products of HTL. Especially for PC, in which 80% of the oil products from HTL without catalyst are readily available as off-the-shelf platform chemicals. For Epoxy, despite also having high-value chemicals present, its oil phase is more complex. This is due to its characteristic branched thermosetting chemical structure, which yields also branched products. In both cases, alkali increases hydrolysis rates yielding more AP products, mainly phenols.

PET: Solid TA is the main product of non-catalytic HTL. The AP is rich in identifiable chemicals and is increased in concentration by addition of KOH during HTL processing. The catalyst also favors decarboxylation of TA, resulting in by products in both solid and AP.

PA6 and PA66: Yields AP monomers that could be used for repolymerization if pure streams are processed. The concentration of such monomers is still of high importance for process feasibility and higher concentrations are desirable. The crystalline phases of these polymers decompose differently and KOH is capable of promoting total depolymerization in short residence times.

PUR: Yields a complex oil, composed of oligomers and a minor part of low boiling point compounds. KOH catalysis tends to increase the small compounds concentration, indicating more hydrolysis of the reactive sites. The small compounds identified and relatively high oil yield, show that PUR is a suitable feedstock for HTL processing.

PVC: HTL processing in the presence of alkali yields increased gas products due to $Cl_2$ release. PVC solid residues are highly dechlorinated, which indicates that this fraction can be further used as carbon source. The dechlorination of carbon in this case is especially valuable as HCl is the main product and is present solely in the AP.

Overall, each type of synthetic polymer presents its own depolymerization characteristics under HTL, which brings opportunities and challenges for future applications in pure and mixed streams. Low residence time subcritical HTL is not able to cope with polyolefins and PS, however it represents a very interesting approach on chemical recycling of heteroatom-containing synthetic polymers.

# 7. ACKNOWLEDGMENTS

This research was funded by the European Union's Horizon 2020 research and innovation program under grant agreement No. 764734 (HyFlexFuel—Hydrothermal liquefaction: Enhanced performance and feedstock flexibility for efficient biofuel production) and The Centre for Circular Bioeconomy (CBIO) of Aarhus University.

# REFERENCES

Anastasakis, K., Biller, P., Madsen, R.B., Glasius, M., Johannsen, I., 2018. Continuous Hydrothermal Liquefaction of Biomass in a Novel Pilot Plant with Heat Recovery and Hydraulic Oscillation. Energies 11, 1–23. https://doi.org/10.3390/en11102695

Arai, R., Zenda, K., Hatakeyama, K., Yui, K., Funazukuri, T., 2010. Reaction kinetics of hydrothermal depolymerization of poly ( ethylene naphthalate ), poly ( ethylene terephthalate ), and polycarbonate with aqueous ammonia solution. Chem. Eng. Sci. 65, 36–41. https://doi.org/10.1016/j.ces.2009.03.023

Bai, B., Jin, H., Fan, C., Cao, C., Wei, W., Cao, W., 2019. Experimental investigation on liquefaction of plastic waste to oil in supercritical water. Waste Manag. https://doi.org/10.1016/j.wasman.2019.04.017

Biller, P., Johannsen, I., dos Passos, J.S., Ottosen, L.D.M., 2018. Primary sewage sludge filtration using biomass filter aids and subsequent hydrothermal co-liquefaction. Water Res. 130, 58–68. https://doi.org/10.1016/j.watres.2017.11.048

Biller, P., Madsen, R.B., Klemmer, M., Becker, J., Iversen, B.B., Glasius, M., 2016. Effect of hydrothermal liquefaction aqueous phase recycling on bio-crude yields and composition. Bioresour. Technol. 220, 190–199. https://doi.org/10.1016/j.biortech.2016.08.053

Brunner, G., 2014. Hydrothermal and Supercritical Water Processes, V 5. ed. Elsevier B.V., Hamburg, Germany.

Castello, D., Haider, M.S., Rosendahl, L.A., 2019. Catalytic upgrading of hydrothermal liquefaction biocrudes: Different challenges for different feedstocks. Renew. Energy 141, 420–430. https://doi.org/10.1016/j.renene.2019.04.003

Castello, D., Pedersen, T.H., 2018. Continuous Hydrothermal Liquefaction of Biomass : A Critical Review. https://doi.org/10.3390/en11113165

Coorporation, R.I., 1982. An investifation of liquefaction of wood. Birmingham, Alabama.

Dai, Z., Hatano, B., Kadokawa, J.I., Tagaya, H., 2002. Effect of diaminotoluene on the decomposition of polyurethane foam waste in superheated water. Polym. Degrad. Stab. 76, 179–184. https://doi.org/10.1016/S0141-3910(02)00010-1

Elliott, D.C., 2011. Hydrothermal Processing, in: Brown, R.C. (Ed.), Thermochemical Processing of Biomass: Conversion into Fuels, Chemicals and Power. John Wiley & Sons Ltd, pp. 200–231. https://doi.org/10.1002/9781119990840.ch7

Foudriaan, F., Peferoen, D.G.R., 1990. Liquid fuels from biomass via a hydrothermal process. Chem. Eng. Sci. 45, 2729–2734.

Ghosh, S.K., Agamuthu, P., 2018. Circular economy : The way forward. https://doi.org/10.1177/0734242X18778444

Hai-feng, Z., Xiao-li, S.U., Dong-kai, S.U.N., Rong, Z., Ji-cheng, B.I., 2007. Investigation on degradation of polyethylene to oil in a continuous supercritical water reactor 35.

Iwaya, T., Sasaki, M., Goto, M., 2006. Kinetic analysis for hydrothermal depolymerization of nylon 6 91, 1989–1995. https://doi.org/10.1016/j.polymdegradstab.2006.02.009

Izzo, B., Harrell, C.L., Klein, M.T., 1997. Nitrile Reaction in High-Temperature Water : Kinetics and Mechanism. AIChE J. 43, 2048–2058.

Izzo, B., Klein, M.T., LaMarca, C., Scrivner, N.C., 1999. Hydrothermal reaction of saturated and unsaturated nitriles: Reactivity and reaction pathway analysis. Ind. Eng. Chem. Res. 38, 1183–1191. https://doi.org/10.1021/ie9803218

Korhonen, J., Honkasalo, A., Seppälä, J., 2018. Circular Economy : The Concept and its Limitations. Ecol. Econ. 143, 37–46. https://doi.org/10.1016/j.ecolecon.2017.06.041

Kwak, H., Shin, H., Bae, S., Kumazawa, H., 2005. Characteristics and Kinetics of Degradation of Polystyrene in Supercritical Water 1–6. https://doi.org/10.1002/app.23896

Madsen, R.B., Jensen, M.M., Mørup, A.J., Houlberg, K., Christensen, P.S., Klemmer, M., Becker, J., Iversen, B.B., Glasius, M., 2016. Using design of experiments to optimize derivatization with methyl chloroformate for quantitative analysis of the aqueous phase from hydrothermal liquefaction of biomass. Anal. Bioanal. Chem. 408, 2171–2183. https://doi.org/10.1007/s00216-016-9321-6

McKeen, L.W., McKeen, L.W., 2012. Polyamides (Nylons). Film Prop. Plast. Elastomers 157–188. https://doi.org/10.1016/B978-1-4557-2551-9.00008-6

Moriya, T., Enomoto, H., 1999. Characteristics of polyethylene cracking in supercritical water compared to thermal cracking. Polym. Degrad. Stab. 65, 373–386. https://doi.org/10.1016/S0141-3910(99)00026-9

Mu, T.E., Hultzsch, K.C., Yus, M., Foubelo, F., Tada, M., 2008. Hydroamination : Direct Addition of Amines to Alkenes and Alkynes 3795–3892.

Park, Y., Hool, J.N., Curtis, C.W., Roberts, C.B., 2001. Depolymerization of Styrene - Butadiene Copolymer in Near-Critical and Supercritical Water 756–767. https://doi.org/10.1021/ie000502l

Pedersen, T.H., Conti, F., 2017. Improving the circular economy via hydrothermal processing of high-density waste plastics Plastic waste Value added. Waste Manag. 68, 24–31. https://doi.org/10.1016/j.wasman.2017.06.002

Sanli, O., 1990. Homogeneous hydrolysis of polyacrylonitrile by potassium hydroxide. Eur. Polym. J. 26, 9–13.

Singh, B., Sharma, N., 2008. Mechanistic implications of plastic degradation 93. https://doi.org/10.1016/j.polymdegradstab.2007.11.008

Skaggs, R.L., Coleman, A.M., Seiple, T.E., Milbrandt, A.R., 2018. Waste-to-Energy biofuel production potential for selected feedstocks in the conterminous United States. Renew. Sustain. Energy Rev. 82, 2640–2651. https://doi.org/10.1016/j.rser.2017.09.107

Su, X., Zhao, Y., Zhang, R., Bi, J., 2004. Investigation on degradation of polyethylene to oils in supercritical water 85, 1249–1258. https://doi.org/10.1016/j.fuproc.2003.11.044

Wan, B., Kao, C., Cheng, W., 2001. Kinetics of Depolymerization of Poly (ethylene terephthalate) in a Potassium Hydroxide Solution. Ind. Eng. Chem 40, 509–514. https://doi.org/10.1021/ie0005304

Wong, S.L., Ngadi, N., Amin, N.A.S., Abdullah, T.A.T., Inuwa, I.M., Ngadi, N., Amin, N.A.S., Abdullah, T.A.T., Inuwa, I.M., 2016. Pyrolysis of low density polyethylene waste in subcritical water optimized by response surface methodology 3330. https://doi.org/10.1080/09593330.2015.1068376

Yildirir, E., Onwudili, J.A., Williams, P.T., 2015. Chemical Recycling of Printed Circuit Board Waste by Depolymerization in Sub- and Supercritical Solvents. Waste and Biomass Valorization 6, 959–965. https://doi.org/10.1007/s12649-015-9426-8

Zenda, K., Funazukuri, T., 2008. Depolymerization of poly ( ethylene terephthalate ) in dilute aqueous ammonia solution under hydrothermal 1386, 1381–1386. https://doi.org/10.1002/jctb

Zhao, X., Xia, Y., Zhan, L., Xie, B., Gao, B., Wang, J., 2018a. Hydrothermal Treatment of E-Waste

Plastics for Tertiary Recycling: Product Slate and Decomposition Mechanisms. ACS Sustain. Chem. Eng. 7, 1464–1473. https://doi.org/10.1021/acssuschemeng.8b05147

Zhao, X., Zhan, L., Xie, B., Gao, B., 2018b. Products derived from waste plastics ( PC , HIPS , ABS , PP and PA6 ) via hydrothermal treatment : Characterization and potential applications. Chemosphere 207, 742–752. https://doi.org/10.1016/j.chemosphere.2018.05.156